\documentclass[conference]{IEEEtran}
\IEEEoverridecommandlockouts
\usepackage{cite}
\usepackage{amsmath,amssymb,amsfonts}
\usepackage{algorithmic}
\usepackage{graphicx}
\usepackage{textcomp}
\usepackage{xcolor}

\usepackage[normalem]{ulem}
\usepackage{cite}
\usepackage{setspace}

\renewcommand\thesection{\arabic{section}}
\renewcommand\thesubsection{\thesection.\arabic{subsection}}
\newcommand\blfootnote[1]{%
  \begingroup
  \renewcommand\thefootnote{}\footnote{#1}%
  \addtocounter{footnote}{-1}%
  \endgroup
}

\newcommand{\disaggrSystems}{DSs}

\newcommand{\disaggrSystem}{DS}
\newcommand{\computeComponent}{CC}
\newcommand{\computeComponents}{CCs}
\newcommand{\memoryComponent}{MC}
\newcommand{\memoryComponents}{MCs}

\newcommand{\myName}{\emph{DaeMon}}

\definecolor{amber}{rgb}{1.0, 0.49, 0.0}
\definecolor{darkgreen}{rgb}{0.0, 0.2, 0.13}
\definecolor{darkbyzantium}{rgb}{0.36, 0.22, 0.33}
\definecolor{darkseagreen}{rgb}{0.56, 0.74, 0.56}
\definecolor{darkspringgreen}{rgb}{0.09, 0.5, 0.27}
\definecolor{dollarbill}{rgb}{0.52, 0.73, 0.4}
\definecolor{MidnightBlue}{RGB}{46,46,208}

\newcommand\cgiannou[1]{\noindent{\color{teal}}} 

\newcommand\revised[1]{\noindent{\color{black}{#1}}} 

\def\BibTeX{{\rm B\kern-.05em{\sc i\kern-.025em b}\kern-.08em
    T\kern-.1667em\lower.7ex\hbox{E}\kern-.125emX}}
\begin{document}

\title{Architectural Support for Efficient Data Movement \vspace{-4pt}\\ in Disaggregated Systems \vspace{-20pt}
}

\newcommand{\affilUofT}[0]{\textsuperscript{\S}}
\newcommand{\affilNTUA}[0]{\textsuperscript{$\dagger$}}
\newcommand{\affilIntel}[0]{\textsuperscript{$\ddagger$}}

\author{
\fontsize{11.4}{8}\selectfont
\parbox[t]{1.02\textwidth}{
\hspace{50pt}
{Christina Giannoula\affilUofT\affilNTUA}\hspace{25pt}
{*Kailong Huang\affilUofT}\hspace{25pt}
{*Jonathan Tang\affilUofT}
{\vspace{4pt}\textcolor{white}{}}
\\
{\vspace{3pt}\textcolor{white}{s}}\hspace{15pt}
{Nectarios Koziris\affilNTUA}\hspace{25pt}
{Georgios Goumas\affilNTUA}\hspace{25pt} 
{Zeshan Chishti\affilIntel}\hspace{25pt}%
{Nandita Vijaykumar\affilUofT} 
\\
\vspace{-7pt}
\centering\small{\emph{{\affilUofT University of Toronto \hspace{10pt} \affilNTUA National Technical University of Athens \hspace{10pt}  \affilIntel Intel Corporation }}}%
}
\vspace{-8pt}
}

\maketitle
\thispagestyle{empty}


\setstretch{0.925}

\section{Data Movement in Disaggregated Systems}

\begin{figure}[h]
    \vspace{-8pt}
    \centering
    \includegraphics[width=0.92\linewidth]{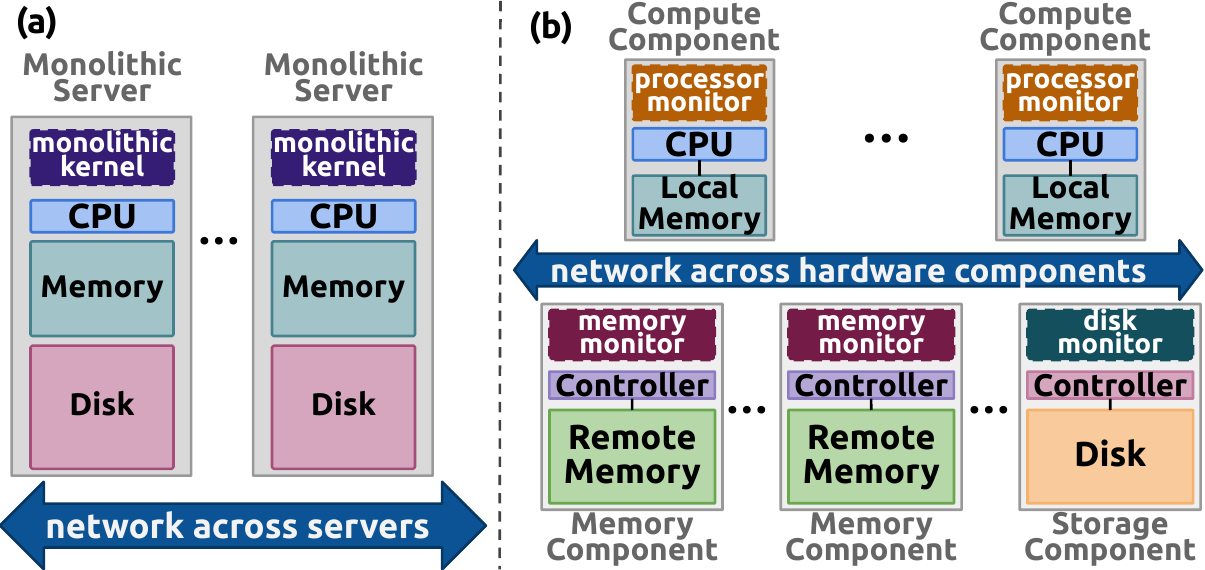}
    \vspace{-4pt}
    \caption{(a) Traditional systems vs (b) \disaggrSystems{}.}
    \label{fig:baseline-architecture-exab}
    \vspace{-7pt}
\end{figure}

\vspace{-2pt}
Traditional data centers include monolithic servers that tightly integrate CPU, memory and disk (Figure~\ref{fig:baseline-architecture-exab}a). Instead, \emph{Disaggregated Systems} (\textbf{\disaggrSystems{}})~\cite{Shan2018LegoOS,guo2021clio,Lee2021MIND} organize multiple compute (\textbf{\computeComponent{}}), memory (\textbf{\memoryComponent{}}) and storage devices as \emph{independent, failure-isolated} components interconnected over a high-bandwidth network (Figure~\ref{fig:baseline-architecture-exab}b). 
\disaggrSystems{} can greatly reduce data center costs by providing improved resource utilization, resource scaling, failure-handling and elasticity in modern data centers~\cite{Calciu2021Rethinking,Shan2018LegoOS,guo2021clio,Lee2021MIND,Giannoula2022Thesis}. \blfootnote{\textbf{* Kailong Huang and Jonathan Tang have equal contribution.}}

\begin{figure}[h]
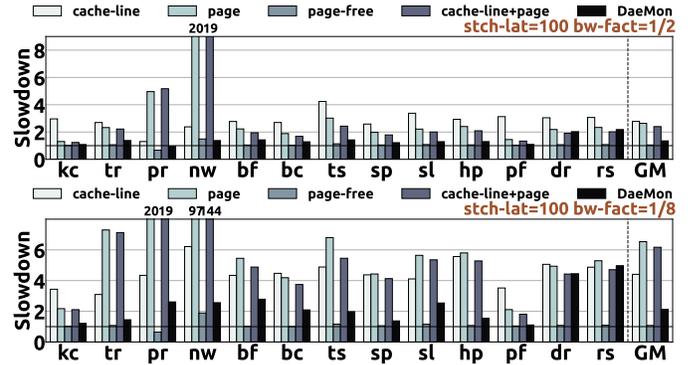

    \vspace{-6pt}
    \centering
    \includegraphics[width=\linewidth]{motivation_exab2_netlat_100_bwsf_2_ipc_slowdown.pdf}
    \includegraphics[width=\linewidth]{motivation_exab2_netlat_100_bwsf_8_ipc_slowdown.pdf}
    \vspace{-16pt}
    \caption{Data movement overheads in \disaggrSystems{}.}
    \label{fig:motivation-exab}
    \vspace{-8pt}
\end{figure}

The \memoryComponents{} provide large pools of main memory (\textbf{\emph{remote memory}}), while the \computeComponents{} include the on-chip caches and a few GBs of DRAM (\textbf{\emph{local memory}}) that acts as a cache of \emph{remote memory}.  In this context, a large fraction of the application's data ($\sim80$\%)~\cite{Shan2018LegoOS,Lee2021MIND,Gao2016Network} is located in \textit{remote memory}, and can cause large performance penalties (see Figure~\ref{fig:motivation-exab}) from remotely accessing data over the network.

Alleviating data access overheads is challenging in \disaggrSystems{} for the following reasons. First, \disaggrSystems{} are not monolithic and comprise independently managed entities: each component has its own hardware controller, and a specialized
kernel monitor uses its own functionality to manage the component it runs on (only communicates with other monitors via network messaging if there is a need to access remote
resources). This characteristic necessitates a distributed and disaggregated solution that can scale to a large number of independent components in the system. Second, there is high variability in remote memory access latencies since they depend on the locations of the \memoryComponents{}, contention from other jobs that share the same network and \memoryComponents{}, and data placements that can vary during runtime or between multiple executions. This necessitates a solution that is robust towards fluctuations in the network/remote memory bandwidth and latencies. Third, a major factor behind the performance slowdowns is the commonly-used approach in \disaggrSystems{}~\cite{Lee2021MIND,Shan2018LegoOS,Calciu2021Rethinking,Gao2016Network} of moving data at page granularity. This approach effectively provides software transparency, low metadata costs in memory management, and high spatial locality in many applications. However, it can cause high bandwidth consumption and network congestion, and often significantly slows down accesses to critical path cache lines in other concurrently accessed pages. 


Figure~\ref{fig:motivation-exab} presents a performance analysis of different data movement schemes using \disaggrSystems{} for a representative set of heterogeneous workloads. We evaluate a \memoryComponent{} and a \computeComponent{} with \emph{local memory} to fit $\sim$20\% of the application data, and the network bandwidth to be 1/2-1/8$\times$ of the bus bandwidth~\cite{Shan2018LegoOS,Gao2016Network}. Performance of all schemes is normalized to that of the monolithic approach where \emph{all} data fits in the \emph{local memory} of the \computeComponent{}. We make three observations.

First, the typically-used \emph{page} scheme of moving data at a page granularity incurs large slowdowns 
over the monolithic \emph{Local} configuration due to transferring large amounts of data over the network that slows down accesses to other pages. Instead, when pages are moved \revised{\emph{for free}}, i.e., \emph{page-free}, performance significantly improves thanks to spatial locality benefits within pages. Second, moving data always at a \emph{fixed} granularity (\textit{cache-line} via LLC or \textit{page} via \emph{local memory}) cannot provide robustness  
across heterogeneous applications and network configurations: some applications can benefit from cache line (e.g., \emph{pr}, \emph{nw}) or page granularity (e.g., \emph{pf}, \emph{dr}) data movements, while the best-performing granularity also depends on network characteristics (e.g., \emph{bf}, \emph{ts}). Third, \emph{naively} moving data at \emph{both} granularities (\textit{cache-line+page}) to serve
data requests with the latency of the packet that arrives earlier to the \computeComponent{} is still quite inefficient, while critical cache line requests are still queued behind large concurrently accessed pages. \revised{In contrast, Figure~\ref{fig:motivation-exab} shows that \myName{} (see §\ref{KeyIdeasbl}) significantly reduces data movement overheads in \disaggrSystems{} across various network/application characteristics.}


\vspace{-1pt}
\section{Prior Work}

Prior works~\cite{Shan2018LegoOS,Angel2020Disaggregation,Nitu2018Zombieland,Han2013Network,Calciu2021Rethinking,Aguilera2017Remote,Zhang2020RethinkingDM,Lim2012System,Chenxi2020Semeru,Pengfei2021OneSided,Bindschaedler2020Hailstorm,Peng2020Underutilization,Lagar2019Software,Gu2017Infiniswap,Aguilera2018Remote, Lee2021MIND,Pinto2020ThymesisFlow,Katrinis2016dredbox,guo2021clio,Buragohain2017DiME,Gao2016Network,Rao2016IsMemory,Zervas2018Optically,Gouk2022Direct,Zhou2022Carbink} propose OS kernels, system-level solutions, software management systems, architectures for \disaggrSystems{}. These works do not tackle the data movement challenge in \disaggrSystems{}, and thus our work is orthogonal to them.

Prior works on hybrid systems~\cite{Dong2010Simple,Liu2017Hardware,Jiang2010CHOP,Kotra2018Chameleon,Prodromou2017MemPod,Agarwal2017Thermostat,Mitesh2015Heterogeneous,Kai2017Unimem,Loh2011Efficiently,Loh2012Supporting,Chou2014Cameo,Ryoo2017Silcfm,Jevdjic2013Footprint} integrate die-stacked DRAM~\cite{HBM} as DRAM cache of a large main memory~\cite{Agarwal2017Thermostat,Dong2010Simple,Jiang2010CHOP} in a monolithic server, and tackle high page movement costs in two-tiered physical memory via page placement/hot page selection schemes or by moving data at smaller granularity, e.g., cache line. However, data movement in \disaggrSystems{} poses fundamentally different challenges.\cgiannou{not addressed in prior works} First, accesses across the network are significantly slower than within the server, thus intelligent page placement/movement cannot by itself address these high costs. Second, \disaggrSystems{} incur significant variations in access latencies 
based on the current network architecture and concurrent jobs sharing the \memoryComponents{}/network, thus necessitating an solution primarily designed for robustness to this variability. Finally, \disaggrSystems{} \revised{include} independently managed \memoryComponents{} and networks \revised{shared by independent \computeComponents{} running unknown jobs.} Thus, unlike hybrid systems, the solution cannot \revised{assume that the memory management at the \memoryComponents{} can be fully controlled by the CPU side. Our work is the first to examine the data movement problem in fully \disaggrSystems{}, and design an effective solution for \disaggrSystems{}.}


\section{\myName{}'s Key Ideas}\label{KeyIdeasbl}

\begin{figure}[h]
    \vspace{-8pt}
    \centering
    \includegraphics[width=\linewidth]{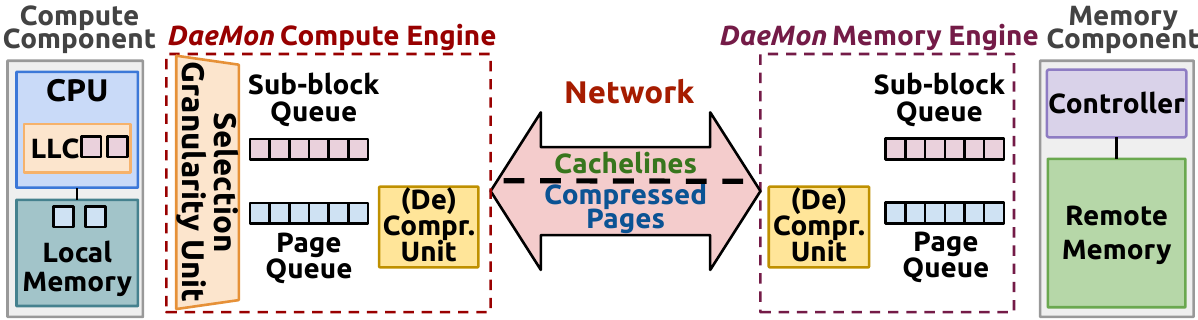}
    \vspace{-16pt}
    \caption{High-level overview of \myName{}.}
    \label{fig:daemon-design-exab}
    \vspace{-2pt}
\end{figure}

\myName{} (Figure~\ref{fig:daemon-design-exab}) is an adaptive and scalable mechanism to alleviate data costs in \disaggrSystems{}, \revised{consisting} of three techniques.

\noindent\textbf{(I) Decoupled Multiple Granularity Data Movement.}  We integrate two separate hardware queues to serve data requests from \emph{remote memory} at two granularities, i.e., cache line (via the \emph{sub-block queue} to LLC) and page (via the \emph{page queue} to \emph{local memory}) granularity, and effectively \textit{prioritize} moving cache lines over moving pages via a \emph{bandwidth partitioning} approach: a queue controller serves cache line requests with a higher \emph{predefined fixed rate} than page requests to ensure that any given time a certain fraction of the bandwidth resources is always allocated to serve cache line moves fast.
This key technique enables (i) low metadata overheads by retaining page migrations, (ii) high performance by leveraging data locality within pages, and (iii) fewer slowdowns in cache line data movements that are on the critical path, from expensive page moves that may have been previously triggered.

\noindent\textbf{(II) Selection Granularity Data Movement.} To provide an adaptive data movement solution, we include in each \computeComponent{} two separate hardware buffers to track pending data migrations for both cache line (via the \emph{inflight sub-block buffer}) and page (via the \emph{inflight page buffer}) granularity, and a selection granularity unit to decide if a data request should be served by cache line, page or \emph{both} based on the utilization of the \emph{inflight} buffers. Given that \myName{} prioritizes cache line over page moves, the inflight buffers are utilized in different ways, allowing us to capture the application behavior and the system load during runtime. For example:
 (a) If there is \emph{low locality} within pages, the page buffer has higher utilization than the sub-block buffer (cache lines are prioritized), thus the selection unit favors moving cache lines and throttles pages (and vice-versa). 
(b) Under low \emph{bandwidth utilization} scenarios, the page buffer utilization is low, thus the selection unit schedules more page movements (or both granularities) to obtain data locality benefits (and vice-versa).

\noindent\textbf{(III) Link Compression on Page Movements.} We employ hardware compression units at both the \computeComponents{} and \memoryComponents{} to highly compress pages migrated over the network. \emph{Link compression} on page moves reduces the network bandwidth consumption and alleviates network bottlenecks.

\noindent\textbf{Synergy of Three Techniques.} 
\myName{} cooperatively integrates all three techniques to significantly alleviate data movement overheads, and provide robustness towards network, architecture and application characteristics: \\
(1) Prioritizing requested cache lines helps \myName{} to tolerate high (de)compression latencies in page migrations over the network, while also leveraging benefits of page migrations (low metadata costs, spatial locality in pages). \\
(2) Compressed pages consume less network bandwidth, enabling \myName{} to reserve part of the bandwidth to effectively prioritize critical path cache line accesses. \\
(3) Compression on page moves helps \myName{} to adapt to the data compressibility: if the pages are highly compressible, the inflight page buffer empties at a faster rate, and \myName{} favors sending more pages (and vice-versa).

\section{\myName{}'s Key Results}

\begin{figure}[h]
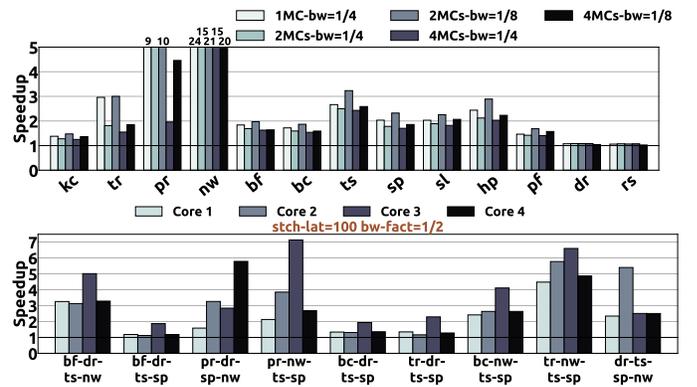

    \vspace{-2pt}
    \centering
    \includegraphics[width=\linewidth]{multimem_exab.pdf}
    \includegraphics[width=\linewidth]{multibench_speedup_netlat_100_bwsf_2_exab.pdf}
    \vspace{-12pt}
    \caption{\myName{}'s benefits over the \emph{page} scheme, (i) varying the \memoryComponents{}, network bandwidth and application, and (ii) running multiple applications in a 4-CPU \computeComponent{} and a \memoryComponent{}.}
    \label{fig:multi-exab}
\end{figure}

We extend Sniper~\cite{carlson2011etloafsaapms} to simulate \disaggrSystems{}, and evaluate capacity intensive workloads from graph processing, HPC, data analytics, bioinformatics, machine learning domains. \myName{} reduces data access costs and improves performance by 3.06$\times$ and 2.39$\times$, respectively, over the widely-adopted approach of moving data at a page granularity. \myName{} leverages the synergy of all three techniques to provide robustness, while retaining the spatial locality, transparency and metadata management benefits of page granularity movements. We show that \myName{} achieves high system performance for various network/architecture configurations and applications (Figure~\ref{fig:multi-exab} top), and multiple concurrently running jobs in the \disaggrSystem{} (Figure~\ref{fig:multi-exab} bottom), compared to the widely-adopted approach of moving data at page granularity.

\pagebreak
\bibliographystyle{unsrt}
\bibliography{references}

\begin{thebibliography}{10}

\bibitem{Shan2018LegoOS}
Yizhou Shan, Yutong Huang, Yilun Chen, and Yiying Zhang.
\newblock {LegoOS: A Disseminated, Distributed OS for Hardware Resource
  Disaggregation}.
\newblock In {\em OSDI}, 2018.

\bibitem{guo2021clio}
Zhiyuan Guo, Yizhou Shan, Xuhao Luo, Yutong Huang, and Yiying Zhang.
\newblock {Clio: A Hardware-Software Co-Designed Disaggregated Memory System}.
\newblock In {\em ASPLOS}, 2022.

\bibitem{Lee2021MIND}
Seung-seob Lee, Yanpeng Yu, Yupeng Tang, Anurag Khandelwal, Lin Zhong, and
  Abhishek Bhattacharjee.
\newblock {MIND: In-Network Memory Management for Disaggregated Data Centers}.
\newblock In {\em SOSP}, 2021.

\bibitem{Calciu2021Rethinking}
Irina Calciu, M.~Talha Imran, Ivan Puddu, Sanidhya Kashyap, Hasan~Al Maruf,
  Onur Mutlu, and Aasheesh Kolli.
\newblock {Rethinking Software Runtimes for Disaggregated Memory}.
\newblock In {\em ASPLOS}, 2021.

\bibitem{Giannoula2022Thesis}
Christina Giannoula.
\newblock {Accelerating Irregular Applications via Efficient Synchronization
  and Data Access Techniques}.
\newblock In {\em arXiv}, 2022.

\bibitem{Gao2016Network}
Peter~X. Gao, Akshay Narayan, Sagar Karandikar, Joao Carreira, Sangjin Han,
  Rachit Agarwal, Sylvia Ratnasamy, and Scott Shenker.
\newblock {Network Requirements for Resource Disaggregation}.
\newblock In {\em OSDI}, 2016.

\bibitem{Angel2020Disaggregation}
Sebastian Angel, Mihir Nanavati, and Siddhartha Sen.
\newblock {Disaggregation and the Application}.
\newblock In {\em HotCloud}, 2020.

\bibitem{Nitu2018Zombieland}
Vlad Nitu, Boris Teabe, Alain Tchana, Canturk Isci, and Daniel Hagimont.
\newblock {Welcome to Zombieland: Practical and Energy-Efficient Memory
  Disaggregation in a Datacenter}.
\newblock In {\em EuroSys}, 2018.

\bibitem{Han2013Network}
Sangjin Han, Norbert Egi, Aurojit Panda, Sylvia Ratnasamy, Guangyu Shi, and
  Scott Shenker.
\newblock {Network Support for Resource Disaggregation in Next-Generation
  Datacenters}.
\newblock In {\em HotNets}, 2013.

\bibitem{Aguilera2017Remote}
Marcos~K. Aguilera, Nadav Amit, Irina Calciu, Xavier Deguillard, Jayneel
  Gandhi, Pratap Subrahmanyam, Lalith Suresh, Kiran Tati, Rajesh
  Venkatasubramanian, and Michael Wei.
\newblock {Remote Memory in the Age of Fast Networks}.
\newblock In {\em SoCC}, 2017.

\bibitem{Zhang2020RethinkingDM}
Qizhen Zhang, Yifan Cai, Sebastian~G. Angel, Vincent Liu, Ang Chen, and B.~T.
  Loo.
\newblock {Rethinking Data Management Systems for Disaggregated Data Centers}.
\newblock In {\em CIDR}, 2020.

\bibitem{Lim2012System}
Kevin Lim, Yoshio Turner, Jose~Renato Santos, Alvin AuYoung, Jichuan Chang,
  Parthasarathy Ranganathan, and Thomas~F. Wenisch.
\newblock {System-Level Implications of Disaggregated Memory}.
\newblock In {\em HPCA}, 2012.

\bibitem{Chenxi2020Semeru}
Chenxi Wang, Haoran Ma, Shi Liu, Yuanqi Li, Zhenyuan Ruan, Khanh Nguyen,
  Michael~D. Bond, Ravi Netravali, Miryung Kim, and Guoqing~Harry Xu.
\newblock {Semeru: A Memory-Disaggregated Managed Runtime}.
\newblock In {\em OSDI}, 2020.

\bibitem{Pengfei2021OneSided}
Pengfei Zuo, Jiazhao Sun, Liu Yang, Shuangwu Zhang, and Yu~Hua.
\newblock {One-sided RDMA-Conscious Extendible Hashing for Disaggregated
  Memory}.
\newblock In {\em ATC}, 2021.

\bibitem{Bindschaedler2020Hailstorm}
Laurent Bindschaedler, Ashvin Goel, and Willy Zwaenepoel.
\newblock {Hailstorm: Disaggregated Compute and Storage for Distributed
  LSM-Based Databases}.
\newblock In {\em ASPLOS}, 2020.

\bibitem{Peng2020Underutilization}
Ivy Peng, Roger Pearce, and Maya Gokhale.
\newblock {On the Memory Underutilization: Exploring Disaggregated Memory on
  HPC Systems}.
\newblock In {\em SBAC-PAD}, 2020.

\bibitem{Lagar2019Software}
Andres Lagar-Cavilla, Junwhan Ahn, Suleiman Souhlal, Neha Agarwal, Radoslaw
  Burny, Shakeel Butt, Jichuan Chang, Ashwin Chaugule, Nan Deng, Junaid Shahid,
  Greg Thelen, Kamil~Adam Yurtsever, Yu~Zhao, and Parthasarathy Ranganathan.
\newblock {Software-Defined Far Memory in Warehouse-Scale Computers}.
\newblock In {\em ASPLOS}, 2019.

\bibitem{Gu2017Infiniswap}
Juncheng Gu, Youngmoon Lee, Yiwen Zhang, Mosharaf Chowdhury, and Kang~G. Shin.
\newblock {Efficient Memory Disaggregation with Infiniswap}.
\newblock In {\em NSDI}, 2017.

\bibitem{Aguilera2018Remote}
Marcos~K. Aguilera, Nadav Amit, Irina Calciu, Xavier Deguillard, Jayneel
  Gandhi, Stanko Novakovi{\'c}, Arun Ramanathan, Pratap Subrahmanyam, Lalith
  Suresh, Kiran Tati, Rajesh Venkatasubramanian, and Michael Wei.
\newblock {Remote Regions: A Simple Abstraction for Remote Memory}.
\newblock In {\em ATC}, 2018.

\bibitem{Pinto2020ThymesisFlow}
Christian Pinto, Dimitris Syrivelis, Michele Gazzetti, Panos Koutsovasilis,
  Andrea Reale, Kostas Katrinis, and H.~Peter Hofstee.
\newblock {ThymesisFlow: A Software-Defined, HW/SW co-Designed Interconnect
  Stack for Rack-Scale Memory Disaggregation}.
\newblock In {\em MICRO}, 2020.

\bibitem{Katrinis2016dredbox}
K.~Katrinis, D.~Syrivelis, D.~Pnevmatikatos, G.~Zervas, D.~Theodoropoulos,
  I.~Koutsopoulos, K.~Hasharoni, D.~Raho, C.~Pinto, F.~Espina, S.~Lopez-Buedo,
  Q.~Chen, M.~Nemirovsky, D.~Roca, H.~Klos, and T.~Berends.
\newblock {Rack-Scale Disaggregated Cloud Data Centers: The dReDBox Project
  Vision}.
\newblock In {\em DATE}, 2016.

\bibitem{Buragohain2017DiME}
Dhantu Buragohain, Abhishek Ghogare, Trishal Patel, Mythili Vutukuru, and
  Purushottam Kulkarni.
\newblock {DiME: A Performance Emulator for Disaggregated Memory
  Architectures}.
\newblock In {\em APSys}, 2017.

\bibitem{Rao2016IsMemory}
Pramod~Subba Rao and George Porter.
\newblock {Is Memory Disaggregation Feasible? A Case Study with Spark SQL}.
\newblock In {\em ANCS}, 2016.

\bibitem{Zervas2018Optically}
Georgios Zervas, Hui Yuan, Arsalan Saljoghei, Qianqiao Chen, and Vaibhawa
  Mishra.
\newblock {Optically Disaggregated Data Centers with Minimal Remote Memory
  Latency: Technologies, Architectures, and Resource Allocation}.
\newblock {\em JOCN}, 2018.

\bibitem{Gouk2022Direct}
Donghyun Gouk, Sangwon Lee, Miryeong Kwon, and Myoungsoo Jung.
\newblock {Direct Access, {High-Performance} Memory Disaggregation with
  {DirectCXL}}.
\newblock In {\em ATC}, 2022.

\bibitem{Zhou2022Carbink}
Yang Zhou, Hassan M.~G. Wassel, Sihang Liu, Jiaqi Gao, James Mickens, Minlan
  Yu, Chris Kennelly, Paul Turner, David~E. Culler, Henry~M. Levy, and Amin
  Vahdat.
\newblock {Carbink: {Fault-Tolerant} Far Memory}.
\newblock In {\em OSDI}, 2022.

\bibitem{Dong2010Simple}
Xiangyu Dong, Yuan Xie, Naveen Muralimanohar, and Norman~P. Jouppi.
\newblock {Simple but Effective Heterogeneous Main Memory with On-Chip Memory
  Controller Support}.
\newblock In {\em SC}, 2010.

\bibitem{Liu2017Hardware}
Haikun Liu, Yujie Chen, Xiaofei Liao, Hai Jin, Bingsheng He, Long Zheng, and
  Rentong Guo.
\newblock {Hardware/Software Cooperative Caching for Hybrid DRAM/NVM Memory
  Architectures}.
\newblock In {\em ICS}, 2017.

\bibitem{Jiang2010CHOP}
Xiaowei Jiang, Niti Madan, Li~Zhao, Mike Upton, Ravishankar Iyer, Srihari
  Makineni, Donald Newell, Yan Solihin, and Rajeev Balasubramonian.
\newblock {CHOP: Adaptive Filter-Based DRAM Caching for CMP Server Platforms}.
\newblock In {\em HPCA}, 2010.

\bibitem{Kotra2018Chameleon}
Jagadish~B. Kotra, Haibo Zhang, Alaa~R. Alameldeen, Chris Wilkerson, and
  Mahmut~T. Kandemir.
\newblock {CHAMELEON: A Dynamically Reconfigurable Heterogeneous Memory
  System}.
\newblock In {\em MICRO}, 2018.

\bibitem{Prodromou2017MemPod}
Andreas Prodromou, Mitesh Meswani, Nuwan Jayasena, Gabriel Loh, and Dean~M.
  Tullsen.
\newblock {MemPod: A Clustered Architecture for Efficient and Scalable
  Migration in Flat Address Space Multi-level Memories}.
\newblock In {\em HPCA}, 2017.

\bibitem{Agarwal2017Thermostat}
Neha Agarwal and Thomas~F. Wenisch.
\newblock {Thermostat: Application-Transparent Page Management for Two-Tiered
  Main Memory}.
\newblock In {\em ASPLOS}, 2017.

\bibitem{Mitesh2015Heterogeneous}
Mitesh~R. Meswani, Sergey Blagodurov, David Roberts, John Slice, Mike
  Ignatowski, and Gabriel~H. Loh.
\newblock {Heterogeneous Memory Architectures: A HW/SW Approach for Mixing
  Die-Stacked and Off-Package Memories}.
\newblock In {\em HPCA}, 2015.

\bibitem{Kai2017Unimem}
Kai Wu, Yingchao Huang, and Dong Li.
\newblock {Unimem: Runtime Data Managementon Non-Volatile Memory-Based
  Heterogeneous Main Memory}.
\newblock In {\em SC}, 2017.

\bibitem{Loh2011Efficiently}
Gabriel~H. Loh and Mark~D. Hill.
\newblock {Efficiently Enabling Conventional Block Sizes for Very Large
  Die-Stacked DRAM Caches}.
\newblock In {\em MICRO}, 2011.

\bibitem{Loh2012Supporting}
Gabriel Loh and Mark~D. Hill.
\newblock {Supporting Very Large DRAM Caches with Compound-Access Scheduling
  and MissMap}.
\newblock {\em IEEE Micro}, 2012.

\bibitem{Chou2014Cameo}
Chia~Chen Chou, Aamer Jaleel, and Moinuddin~K. Qureshi.
\newblock {CAMEO: A Two-Level Memory Organization with Capacity of Main Memory
  and Flexibility of Hardware-Managed Cache}.
\newblock In {\em MICRO}, 2014.

\bibitem{Ryoo2017Silcfm}
Jee~Ho Ryoo, Mitesh~R. Meswani, Andreas Prodromou, and Lizy~K. John.
\newblock Silc-fm: Subblocked interleaved cache-like flat memory organization.
\newblock In {\em HPCA}, 2017.

\bibitem{Jevdjic2013Footprint}
Djordje Jevdjic, Stavros Volos, and Babak Falsafi.
\newblock {Die-Stacked DRAM Caches for Servers: Hit Ratio, Latency, or
  Bandwidth? Have It All with Footprint Cache}.
\newblock In {\em ISCA}, 2013.

\bibitem{HBM}
Hongshin Jun, Jinhee Cho, Kangseol Lee, Ho-Young Son, Kwiwook Kim, Hanho Jin,
  and Keith Kim.
\newblock {HBM DRAM Technology and Architecture}.
\newblock In {\em IMW}, 2017.

\bibitem{carlson2011etloafsaapms}
Trevor~E. Carlson, Wim Heirman, and Lieven Eeckhout.
\newblock {Sniper: Exploring the Level of Abstraction for Scalable and Accurate
  Parallel Multi-Core Simulations}.
\newblock In {\em SC}, 2011.

\end{thebibliography}

\end{document}